\documentclass[12pt]{iopart}
\usepackage{graphicx}
\usepackage{gensymb}

\usepackage{iopams} 
\begin{document}

\title[Surface science motivated by heating of trapped ions from the quantum ground state]{Surface science motivated by heating of trapped ions from the quantum ground state}

\author{D A Hite$^1$, K S McKay$^{1,2}$, and D P Pappas$^1$}

\address{$^1$ National Institute of Standards and Technology, Boulder, Colorado 80305 USA}
\address{$^2$ Department of Physics, University of Colorado, Boulder, Colorado 80309 USA}

\ead{dustin.hite@nist.gov}
\vspace{10pt}
%\begin{indented}
%\item[]September 2021
%\end{indented}

\begin{abstract}
For the past two and a half decades, anomalous heating of trapped ions from nearby electrode surfaces has continued to demonstrate unexpected results.  Caused by electric-field noise, this heating of the ions' motional modes remains an obstacle for scalable quantum computation with trapped ions.  One of the anomalous features of this electric-field noise is the reported nonmonotonic behavior in the heating rate when a trap is incrementally cleaned by ion bombardment.  Motivated by this result, the present work reports on a surface analysis of a sample ion-trap electrode treated similarly with incremental doses of Ar$^+$ ion bombardment.  Kelvin probe force microscopy and x-ray photoelectron spectroscopy were used to investigate how the work functions on the electrode surface vary depending on the residual contaminant coverage between each treatment.  It is shown that the as-fabricated Au electrode is covered with a hydrocarbon film that is modified after the first treatment, resulting in work functions and core-level binding energies that resemble that of atomic-like carbon on Au.  Changes in the spatial distribution of work functions with each treatment, combined with a suggested phenomenological coverage and surface-potential roughness dependence to the heating, appear to be related to the nonmonotonic behavior previously reported.
\end{abstract}

\vspace{2pc}

%\submitto{\NJP}
%\maketitle

\section{Introduction}

In electrostatics, metal electrodes are represented as having equipotential surfaces, which is a valid idealization for most applications.  In some cases, however, nonideal surfaces can influence surface-sensitive sensors and measurements.   In these cases, sensitivity to spatial distributions of regions of different work functions (or surface-potential patches) affects a measurement outcome.   These patches can originate from nonuniform composition or a variable coverage of adsorbates on the surface or even from different crystallographic orientations in an otherwise clean polycrystalline electrode.  For example, in Casimir force experiments \cite{Xu2018}, atomic force sensing measurements \cite{Stipe2001}, and space-based gravitational wave detection tests \cite{Pollack2008}, a residual background potential becomes a precision measurement problem when it varies spatially.  In ion-trap systems, surface patch potentials can fluctuate temporally causing electric-field noise at the location and motional frequency of the trapped ions \cite{Wineland1998}.  This broad-band electric-field noise couples to the net charge of the trapped ions causing the motional modes of the ions in the trap to acquire energy.  This motional heating is an obstacle to scalability in trapped-ion quantum information processing (QIP) because absorption of one phonon during a quantum gate operation ruins the fidelity of the computation.

Exactly which microscopic mechanisms play roles in electric-field noise from surfaces remains unknown and their determination is an active area of experimental and theoretical research.  Various models for the origin of this noise have been put forth over nearly two decades, but none have been able to fully describe all of the experimental observations \cite{Brownnutt2015}.  However, the manner in which the noise spectral density $S(f)$ scales with ion-electrode separation $d$ \cite{Turchette2000, Deslauriers2006, Hite2017, Boldin2018, Sedlacek2018}, electrode temperature $T$ \cite{Deslauriers2006, Labaziewicz2008, Chiaverini2014, Bruzewicz2015, Sedlacek2018, Noel2019, Berlin-Udi2021}, frequency $f$ \cite{Turchette2000, Deslauriers2006, Labaziewicz2008, Allcock2011, Hite2012, Daniilidis2014, Bruzewicz2015, Sedlacek2018}, superconducting-electrode phase transition \cite{Wang2010, Chiaverini2014}, degree of adsorbate coverage and surface condition \cite{Allcock2011, Hite2012, Daniilidis2014, McKay2014, Kim2017, McConnell2015, Sedlacek2018a}, and electrode material composition \cite{Hite2013, Chiaverini2014}, all point to processes at the surface as the origin of the noise.  More than one process may contribute to the noise at any one time, but not many surface processes would give rise to this type of electric-field noise.  For example, local work functions of electrode surfaces vary with adsorbate coverage \cite{Jooya2018}. If the coverage was driven to fluctuate temporally, electric-field noise would be present at the location of the ion.  Another proposed mechanism is that of vibrationally fluctuating stationary adsorbate dipoles \cite{Safavi-Naini2011, Safavi-Naini2013, Ray2019}.  Moreover, when surface diffusion of adsorbate dipoles is considered, there is a greater range of electric-field noise due to the atomic site dependence on the electric dipole moment \cite{Kim2017}.  Another model to be considered derives electric-field noise from dielectric loss in overlayers of contamination on the electrode surface \cite{Kumph2016}. Recent experiments have shown that the bulk effects of a dielectric material positioned near a trapped ion cause motional heating that fits well to a dielectric model with no free parameters \cite{Teller2021}.  Teller et al. extrapolate to a thin film and estimate significant
electric-field noise spectral densities that can potentially hinder scalable quantum computation with trapped ions.

The electric-field noise spectral density tends to have a $1⁄f^{\alpha}$-type spectrum; however, experimentally, a wide range of exponents $\alpha$ has been reported in the approximate range 0.5 $<$ $\alpha$ $<$ 2 \cite{Brownnutt2015}.  By measuring the heating rate $\dot{\bar{n}}$ of a ground-state cooled trapped ion as a function of the motional-mode frequency \textit{f}, one can determine the electric-field noise spectral density $S_E(f)$, making use of the relation \cite{Turchette2000}:

\begin{equation}
    S_E(f) = \frac{4mhf}{q^2}\dot{\bar{n}},
\end{equation}

where \textit{m} and \textit{q} are the mass and charge of the ion, and \textit{h} is the Planck constant.
Since $S_E \sim 1/f^{\alpha}$ , then from (1), the frequency scaling of the measured heating rate goes as $\dot{\bar{n}} \sim 1/f^{\alpha +1}$.  Ideally, the frequency scaling of the noise would indicate the type of responsible mechanism.  The noise has been attributed to fluctuations in adsorbate contamination; however, how the adsorbates respond to different driving forces remains unknown.  Typically, when motional heating rates are measured as a function of some independent variable, e.g., \textit{d}, \textit{T}, or \textit{f}, the behavior is monotonic.  That is, as \textit{d} or \textit{f} are decreased, the heating rate increases; as \textit{T} is decreased, the heating rate decreases.  Even in some ion traps that have undergone in-situ surface treatments, such as ion bombardment or laser cleaning, heating rates tend to be reduced by orders of magnitude with a large enough dose.  To understand how motional heating rates depend on the degree of adsorbate coverage, a study was recently reported where an as-fabricated ion-trap electrode surface was treated incrementally by ion bombardment to step-wise reduce the coverage, followed by heating-rate measurements for each residual coverage \cite{Kim2017}.  In that work, a nonmonotonic behavior was observed as the coverage was decreased.  These data provide an important clue to understanding the driving mechanism and response of surface processes giving rise to this anomalous electric-field noise.
	
In this article, variations in the frequency scaling as a function of adsorbate coverage are presented, as the adsorbate contamination layer is incrementally removed by Ar$^+$ ion bombardment in situ.  These data add to the single-frequency data presented in \cite{Kim2017}, where the electric-field-noise behavior is nonmonotonic dependent on the ion-bombardment dose.  The frequency dependence of the heating rates for a given coverage follows a power-law relation over the range of accessible frequencies for this trap.  The variation in the power-law exponent $\alpha$ indicates that there may be a single \cite{Ray2019} or even multiple mechanisms \cite{Foulon2021} arising from surface treatments or a change in coverage.  Heating rates also are shown to have anisotropic values for the two orthogonal motional modes parallel to the surface of the trap electrodes.  Surface analysis conducted on a sample electrode (not used for the heating-rate measurements) is then presented and compared to the model system C/Au(110) \cite{Jooya2018} because carbon is a ubiquitous contaminant on Au, commonly used for ion-trap electrodes.  Work-function and core-level measurements show that the initial contamination layer on the as-fabricated surface-analysis sample electrode is fundamentally different from atomic carbon deposited on Au, however after the first ion-bombardment treatment, the average work function and core-level binding energies of the residual overlayer resemble that of C/Au(110).  Finally, histograms of Kelvin probe force microscopy (KPFM) images are presented to elucidate the surface-potential “roughness” of the incrementally sputtered electroplated gold electrodes, in contrast to the smooth surface-potential landscapes of deposited carbon on Au(110), quantifying work-function patch sizes with wide distributions of work functions on the electrode surface.  These resulting work-function distributions are not fully explained by the different crystalline orientations within the polycrystalline electrode \cite{Jooya2019} but may also be the result of morphological effects of a sputtered and unannealed surface.  Finally, a simple model is presented to qualitatively explain a hypothesized correlation between the nonmonotonic heating rates and the roughness of the surface-potential maps.

\section{Experimental Details}

The experiments were carried out in two separate ultra-high vacuum chambers ($\sim$ 2 $\times$ $10^{-8}$ Pa base pressure).  The ion-trap chamber was equipped with a cylindrical mirror electron energy analyzer for Auger electron spectroscopy (AES) of the trap-electrode surfaces and Ar$^+$ ion bombardment facilities including gas handling, valves, a turbomolecular pump, and a 500 V to 2 kV ion source for treatments to the trap electrodes.  The surface-analysis chamber was equipped with a hemispherical electron energy analyzer for x-ray photoelectron spectroscopy (XPS) and an atomic force microscope (AFM) for KPFM with similar ion bombardment facilities.  All measurements were conducted at room temperature.

The surfaces of three different samples were studied in this work:  1) the stylus ion-trap electrodes used for heating-rate measurements, 2) an electroplated sample ion-trap electrode used in the KPFM/XPS studies, and 3) a Au(110) single crystal with deposited carbon used as a model system.  The as-fabricated stylus electrodes and sample ion-trap electrode were sputter-treated incrementally to study the removal of the carbonaceous overlayer.  For the C/Au(110) sample, atomic carbon was deposited by sublimation onto a cleaned Au(110) surface as detailed in \cite{Jooya2018}. 

\subsection{Heating Rate Measurements}

The electroplated-Au ion-trap electrodes for the heating-rate measurements were designed at NIST and microfabricated by Sandia National Laboratories using synchrotron-based x-ray lithography techniques, commonly used to form high-aspect ratio microstructures \cite{Malek2004}.  These electrodes have a stylus-type geometry with $\sim$ 60 $\mu$m diameter pillars, extending 180 $\mu$m above a 5 mm $\times$ 5 mm substrate with interconnect traces and a large ground plane (figure 1).  The original macroscopic stylus-type radio-frequency (RF) Paul trap was described by Maiwald et al. \cite{Maiwald2009}, and previous design iterations of the microfabricated version have been described by Arrington et al. \cite{Arrington2013} and McKay et al. \cite{McKay2014}.

The trap used in this work is the same as that used by Kim et al. \cite{Kim2017}, wherein individual $^{25}$Mg$^+$ ions are confined approximately 63 $\mu$m above the center electrode (figure 1).  Heating-rate data are collected for both ground-state cooled radial modes (two orthogonal motional modes oriented in a plane nearly parallel to the electrode surfaces, see figure 1) for a range of motional frequencies at each step of incremental treatments by Ar$^+$ ion bombardment for step-wise removal of the contamination layer.  A simplified description of the heating-rate measurement technique can be found in \cite{Hite2013} adapted from the in-depth description in \cite{Wineland1998}.

\begin{figure}
\centering
\includegraphics[width=0.9 \textwidth, keepaspectratio]{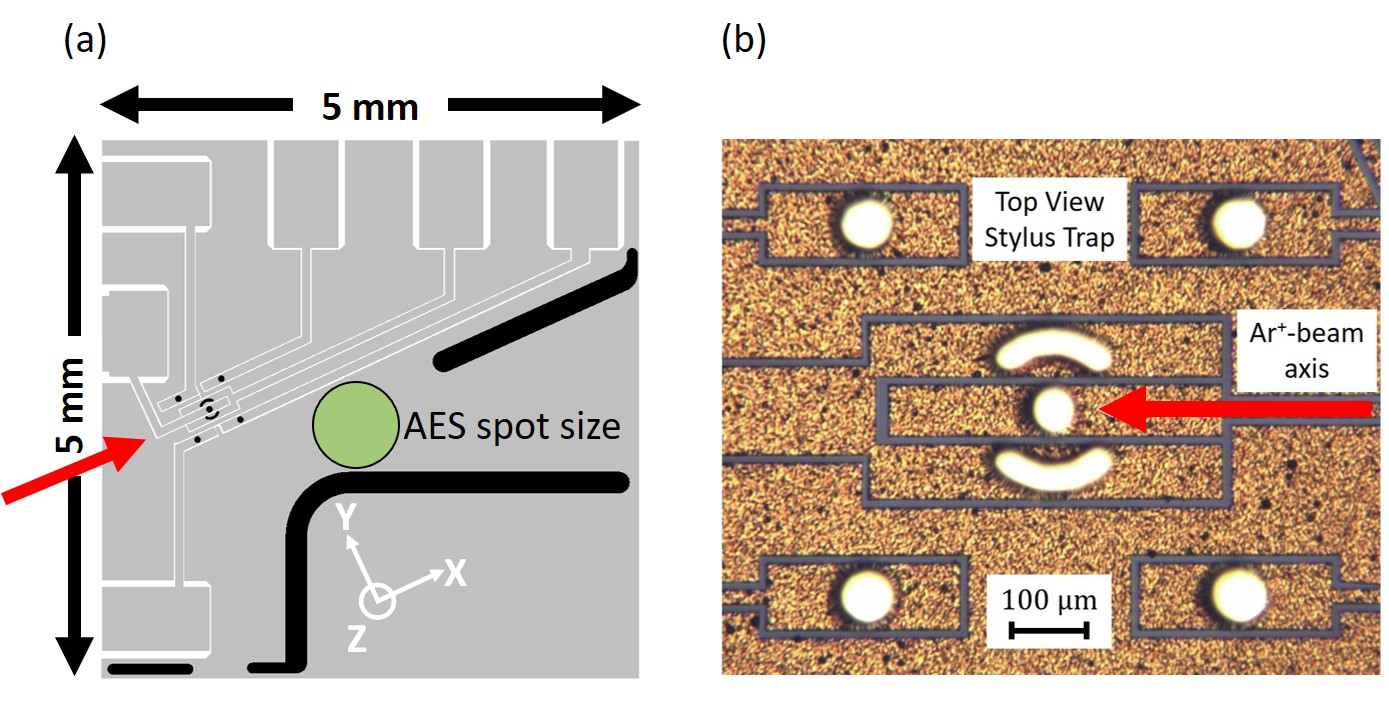}
\caption{(a) Top-view layout schematic of a stylus ion trap chip showing the motional mode axes, Ar$^+$ ion beam direction (red arrow), and Auger beam spot size and location (green circle), $\sim$ 1 mm from the trapping center.  Other features are described in \cite{Arrington2013}.  (b) Top-view optical image of the trapping center.  The light-colored arcs comprise the RF electrode and the 5 circular electrodes ($\sim$ 60 $\mu$m diameter) are DC control electrodes at RF ground.  The electrodes are extruded $\sim$ 180 $\mu$m above a 5 mm $\times$ 5 mm ground plane (gold color).  $^{25}$Mg$^+$ is trapped 63 $\mu$m above the center electrode.  The radial motional modes are in a plane nearly parallel to the top electrode surfaces and the Ar$^+$ beam is oriented at a 30$\degree$ angle with respect to the surface normal.}
\label{fig:Trap}
\end{figure}

The motional-mode frequency is varied by changing the RF potential applied to the trap electrodes.  Five static-potential electrodes are used for adjusting the position of the ion in the null of the RF field and are not used to change the motional frequencies of the ion.  Therefore, changes in the applied RF power are made to vary the motional-mode frequencies.  To confirm a flat response of the heating rate due to various magnitudes of applied RF power, three different RF-drive resonators (50, 65, 93 MHz resonance frequencies) were employed to acquire heating-rate measurements with different RF power, but with the same range of motional secular frequencies.  Since the \textit{x} and \textit{y} motional modes display different frequency scalings, measured to be $\alpha_x$ = 1.6 and $\alpha_y$ = 1.1, a frequency-normalized electric-field noise spectral density, $f^{\alpha}S_E$, was used to correct for the different mode frequencies ranging from approximately 2 MHz to 7 MHz.  The noise amplitude varied by less than a factor of two over an RF power range of approximately 8 dB.  This confirms the flat response of heating rates to applied RF power over the operational range in this ion trap.

The surface contamination level, or the degree of carbon coverage, was measured in situ by AES directly at the trap center (electron beam diameter was $\sim$ 800 $\mu$m) before the first and after the last cleaning treatments to minimize the possibility of adding contamination from the AES electron gun to the trap electrodes nearest the ion.  All other coverage measurements after each of the incremental treatments were made at a location about 1 mm away from the trap electrodes nearest the ion.  Therefore, coverages for the heating rate measurements are reported to be estimated for the surfaces that most affect the ion’s motion.  The estimates are based on depth profiling studies of contamination removal in a different chamber from similar air-exposed electroplated Au sample ion-trap electrodes, using the Ar$^+$ ion-beam dose as the referenced quantity.  Therefore, there is considerable uncertainty in the reported estimated coverage for the intermediate data points on the order of 50\% ($\pm$ 0.2 ML minimum).

The Ar$^+$ ion beam was positioned at a 30$\degree$ angle with respect to the normal to the trap surface and directed along the \textit{x} axis of the trap (see figure 1).  Due to the three-dimensional structure of the stylus trap electrodes, some regions of the electrodes' sidewalls are shadowed from the Ar$^+$ ion beam, which could leave regions near the Mg$^+$ ion untreated ($\gtrsim$ 70 $\mu$m distance to the ion), adding to the uncertainty in the estimated coverage.  Typical parameters for the incremental sputter treatments were 2 kV beam voltage, 0.5 $\mu$A/cm$^2$ sample-current density at 6 $\times$ 10$^{-3}$ Pa Ar for 5 min to 10 min each.  To study the removal of the native contamination on the electrodes, a pre-assembly treatment, prescribed in \cite{McKay2014}, was not conducted in this work.

\subsection{Surface Analysis Measurements}
The sample electrode for surface analysis was made of 8 $\mu$m thick electroplated Au, microfabricated on a sapphire substrate, depicted in \cite{Burd2020, Srinivas2020} and acquired from the same wafer as the linear surface-electrode trap used for quantum gate experiments described in \cite{Srinivas2019, Burd2019}. These electrodes and those used for the heating-rate measurements are nominally the same as they are both air-exposed, initially untreated after fabrication, and composed of polycrystalline electroplated Au.  Differences in fabrication features between the samples, like film thickness and interelectrode gap widths, have no consequence for the surface analysis measurements, except perhaps for the two AES measurements at the trap center, however no evidence of the substrate was observed due to the narrow gaps.  In addition, scanning tunneling microscopy data collected from a nearly duplicate stylus trap showed similar morphologies to the sample electrode discussed in this work before and after ion-bombardment treatments.  Nevertheless, there are limitations to comparing two different surfaces that are initially prepared in an uncontrolled manner, i.e., final steps of cleaning with solvents and air exposure.  As in the heating-rate experiment, the surface-analysis sample is incrementally treated with Ar$^+$ ion bombardment.  The dose for each treatment was approximately 0.3 J/cm$^2$.  Between each treatment, spatially resolved local work functions and residual coverages were measured with KPFM and XPS, respectively.  This allows for the observation of work-function magnitudes, patch sizes, and spatial distributions as a function of the residual coverage.

For the work function measurements, a Cr/Pt-coated Si cantilever was employed in frequency-modulated KPFM to measure the contact potential difference (CPD) between the tip of the cantilever and the sample, typically over a 300 nm $\times$ 300 nm area.  After each CPD measurement, the work function of the tip was recalibrated using a cleaved highly oriented pyrolytic graphite (HOPG) reference sample.  The work function of the HOPG is taken to be 4.6 eV \cite{Takahashi1985}.  This accounts for tip changes that may occur and allows for the relative determination of the sample work function with a spatial resolution of $\sim$ 5 nm to 10 nm.  Values for the sample work function for a given imaged area are reported as the work function for which the image histogram has a peak(s), fitted with a Gaussian distribution(s) with typical widths of approximately 30 meV full-width at half-maximum.

The coverage and core-level binding energies were determined by XPS.  For these measurements, x-rays from a Mg anode were used, producing photons at 1253.6 eV.  To calibrate the binding energy, the work function of the electron energy analyzer was determined using a clean Pd sample providing a large density of states at the Fermi edge.  The energy calibration was performed for two photon energies, 1253.6 and 1486.7 eV.  The coverage is estimated from a simple overlayer model that takes into account the ratios of the peak intensities (for C and Au, in this case) and the exponential attenuation of electrons from depths within the sample using the inelastic mean free path for the particular electron kinetic energies \cite{Seah1979}.  Due to the limitations of this simple overlayer model and the averaged signal over a 200 $\mu$m diameter XPS-analysis area, the uncertainty in the coverage measurement is estimated to be approximately $\pm$  0.2 monolayers (ML) to account for inhomogeneities on the sample surfaces.

\section{Results}

\subsection{Heating Rate Measurements: Frequency Scalings}

Heating rates for several motional mode frequencies in this stylus trap were measured. Figure 2(a) shows the heating rate at 4.7 MHz as a function of the estimated coverage for the electrode surfaces closest to the trapped ion.  The electric-field noise is dominated by the closest electrode surfaces due to the $\sim$ 1/$d^3$ distance dependence \cite{Hite2017}.  These data are the same as those reported in \cite{Kim2017}. The nonmonotonic behavior shows an order of magnitude increase in the submonolayer regime ($\sim$ 0.5 ML), followed by a nearly two orders of magnitude decrease as the coverage $\theta$ is reduced (by incremental sputter treatments) from 3 ML toward a nominal coverage of 0 ML.

%\begin{figure*}[h!]
%\includegraphics[width=0.7 \textwidth, %keepaspectratio]{Fig2ab.jpg}
%\caption{Kelvin images and histograms for C/Au(110) and %sample trap electrode.}
%\label{fig:Trap}
%\end{figure*}

\begin{figure}
\centering
\includegraphics[width=0.6 \textwidth, keepaspectratio]{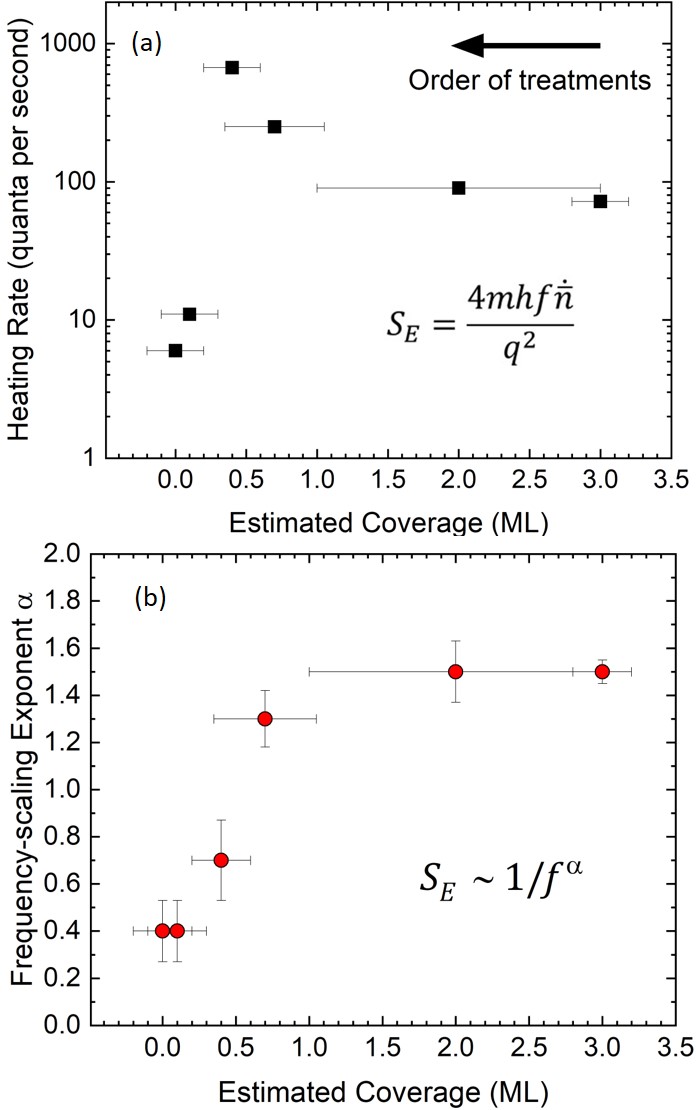}
\caption{(a) Heating rates at a 4.7 MHz motional frequency and (b) frequency-scaling exponents $\alpha$ for an incrementally cleaned ion trap as a function of the estimated residual carbon coverage fit for both directional modes.  The data were collected from high to low coverage after each incremental treatment.  The nonmonotonic dependence on the heating rate was previously reported in \cite{Kim2017}.  Approximately 50\% uncertainty (0.2 ML minimum uncertainty) in the degree of carbon coverage is due to the indirect measurement with AES at a position 1 mm from the trap electrode nearest the ion.}
\label{fig:HeatingRate}
\end{figure}

For various motional-mode frequencies, the heating-rate scaling, characterized by the 1/$f^{\alpha}$ exponent ${\alpha}$, also exhibits a correlated dependence, shown in figure 2(b).  The exponent ${\alpha}$ from fitting to both modes is changed from 1.5 toward smaller values ($\sim$ 0.4) as the coverage is reduced from 0.5 ML toward 0 ML (cleaned surface).  Theoretical models of independently fluctuating adsorbate dipoles have predicted electric-field noise spectra that exhibit the characteristic 1/$f^{\alpha}$ behavior and also show changes from $\alpha$ $\sim$ 1 to $\alpha$ $\sim$ 0 \cite{Safavi-Naini2013, Ray2019}.  Interestingly, in \cite{Ray2019}, it is shown that, for a given frequency, if the adsorbate species is changed, for example, by way of hydrocarbon decomposition, then the spectrum can change from 1/$f^{\alpha}$ at a higher noise level to white noise at a lower noise level.  However, there are discrepancies in magnitude and frequency ranges compared with experimental data that led Ray et al. to suggest that additional processes need to be considered to explain observed heating rates \cite{Ray2019}.  Recently, a study of the collective motion of adsorbates with various mechanisms in different coverage regimes found 1/\textit{f}-type noise extending down into the MHz range with behaviors similar to that report here.  While not as pronounced as the changes seen in figure 2(b),  other work on in-situ surface-treated ion traps also reported decreased scaling exponents \cite{Daniilidis2014, Allcock2011}, however another work did not show a significant change in ${\alpha}$ after achieving low noise levels \cite{Hite2012}.  In a study of \textit{ex situ} sputter-treated electrodes, i.e., exposed to air after treatment, ${\alpha}$ also was not seen to change from before and then after the treatments \cite{Sedlacek2018a}.

In \cite{Schindler2015}, the polarization of electric-field noise from surfaces was investigated to discern between surface noise and technical noise.  In that work, Schindler et al. describe a model to determine the asymmetry between two orthogonal motional modes rotated through an angle $\gamma$ with respect to the surface normal.  At $\gamma$ = 0, one mode was perpendicular to the trap surface (\textit{z}-direction).  Their model predicts a heating-rate ratio \textit{R} =$\dot{\bar{n}}_z$/$\dot{\bar{n}}_x$ for the two modes to be $R = 2$.  Heating rates were measured as a function of the rotation angle and the ratio followed the trend of the model, but with a maximum $R = 4.2$ \cite{Schindler2015}.  

In this work, a distinct asymmetry in heating rates was measured for the two orthogonal modes in a plane nearly parallel to the electrode surface after the fourth treatment (dosages of the 5 total treatments are reported in \cite{Kim2017}).  For the second to last treatment, at a motional mode frequency of $\sim$ 3 MHz, the heating-rate ratio (\textit{R}  = $\dot{\bar{n}}_x$/$\dot{\bar{n}}_y$) was measured to be $R = 3.5$, shown in figure 3.  An anisotropy in the scaling is observed between the two radial modes before treatments (at $\sim$ 3 MHz, $\alpha_x$ = 1.6 and $\alpha_y$ = 1.1, \textit{R} = 1.04), but the directionality in heating rates becomes more pronounced after the incremental sputter treatments and in the low-coverage regime ($\theta$ $\leq$ 0.5 ML).

\begin{figure}
\centering
\includegraphics[width=0.6 \textwidth, keepaspectratio]{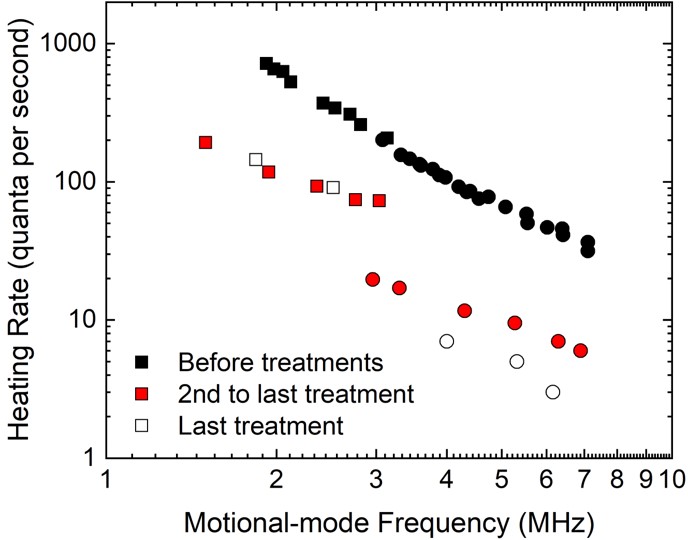}
\caption{Heating rates vs. motional frequency for the \textit{x}- (squares) and \textit{y}-direction (circles) radial modes before and after sputter treatments (5 total treatments).  Anisotropic heating rates become more pronounced in the low-coverage regime ($<$ 0.5 ML).  The mode parallel to the sputter beam projection (\textit{x}) experiences higher heating rates.  For the second to last treatment, at 3 MHz, the asymmetry heating-rate ratio \textit{R} (= $\dot{\bar{n}}_x$/$\dot{\bar{n}}_y$) = 3.5.  Error bars are smaller than the data point symbols.}
\label{fig:frequency}
\end{figure}

Dipolar noise sources in ion traps were also investigated theoretically in the stylus-trap geometry by Galve et al. \cite{Galve2017}.  There, it is pointed out that various dipole orientations can cause different noise levels and in different directions.  The authors suggested that due to the directionality of a sputter beam, nanochannels generated on the Au electrode may play a role in the preferential orientation of surface noise sources.  A rippled surface morphology is common to sputtered and unannealed metal surfaces \cite{Bradley1988}, where the ridges and troughs are oriented perpendicular to the angled (30$\degree$ in this case) sputter-beam direction.  While the spatial resolution of KPFM in this work was not high enough to observe nanochannels in the sample ion-trap electrode, linescans from the corresponding topography from the higher resolution AFM signal revealed a 4 nm periodicity in the \textit{x}-direction over a 50 nm diameter grain.  No periodicity was observed in the \textit{y}-direction.  Considering the relative orientation between the motional mode directions and that of the sputter beam (figure 1), higher heating rates were measured in the direction parallel to the planar projection of the sputter beam.  Additional experimental work is needed to further elucidate this effect.

Noise from technical sources, i.e., other than surface related noise, is a concern especially when heating from the surface has been reduced by an order of magnitude by ion bombardment, for example.  Tests for technical noise sources, such as Johnson noise from circuitry, voltage sources, environmental pickup, and stray resonant light, were routinely conducted in this work.  Details of these tests can be found in Appendix A of \cite{McKay2021}.  Calculations of Johnson noise are orders of magnitude lower than the noise inferred from the lowest measured heating rates in this trap.  Therefore, technical noise is ruled out as the cause for the changed exponent $\alpha$ and the asymmetry in heating in the two motional modes.

\subsection{Surface Analysis of a Sample Ion-trap Electrode: Sputter-treated Electroplated Au}

The fluctuation of work-function patches on electrode surfaces is a mechanism thought to play a role in motional heating of trapped ions \cite{Turchette2000}.  Also, it has been shown that the work function of a Au(110) crystal is strongly dependent on the coverage of deposited atomic carbon \cite{Jooya2018}.  In the submonolayer regime, the measured work function varies by approximately 0.8 eV.  If a dynamic driving mechanism causes the coverage of various work function patches to fluctuate spatially and temporally, then electric-field noise is generated above the surface of an electrode.  To investigate this effect in relation to the nonmonotonic heating-rate dependence shown in figure 2(a), incrementally sputtered ion-trap sample electrodes were studied with static KPFM.  

\begin{figure}
\centering
\includegraphics[width=0.6 \textwidth, keepaspectratio]{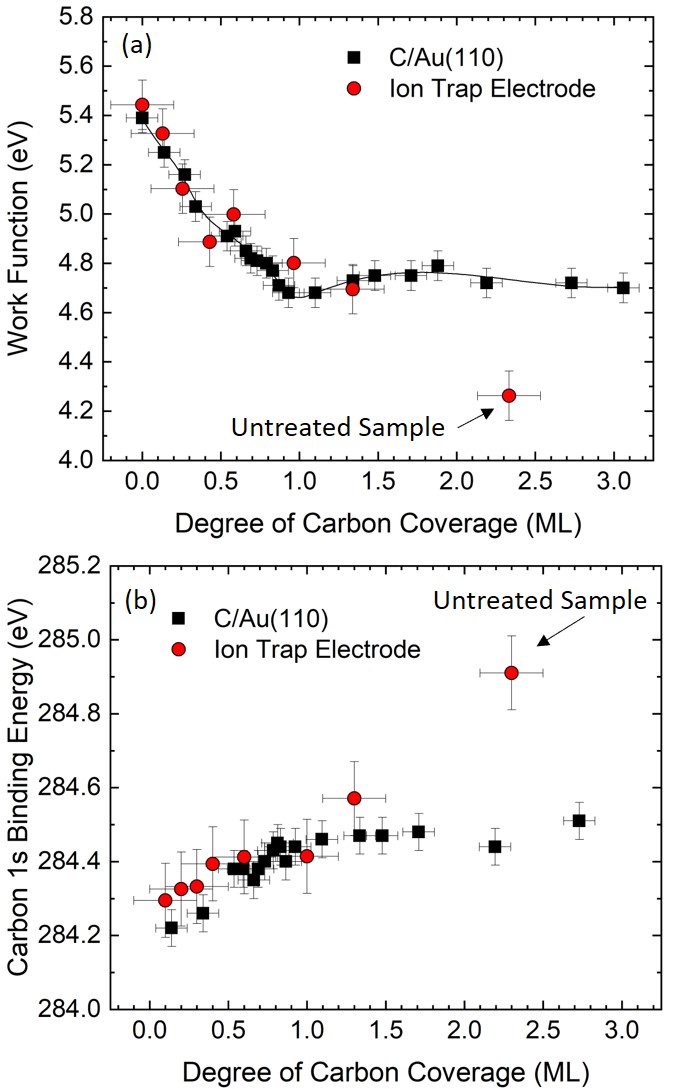}
\caption{(a) Work functions and (b) C 1s binding energies vs. carbon coverage for a sample ion-trap electrode and the model system C/Au(110) for comparison \cite{Jooya2018}.  For the sample ion-trap electrode (red circles), the data were collected from high to low coverage after each incremental treatment.  These data indicate that the as-fabricated sample-electrode surface is composed of a 2.3 ML hydrocarbon film on Au that becomes modified after the first sputter treatment to resemble the surface properties of atomic carbon on Au(110).  The solid line in (a) is a spline fit to DFT-calculated data for C/Au(110) \cite{Jooya2018}.}
\label{fig:WFcorelevels}
\end{figure}

Work functions for an ion-trap electrode are shown in figure 4(a), along with the results of C/Au(110) from \cite{Jooya2018} for comparison.  Initially, the as-fabricated trap electrode is covered with approximately 2.3 ML of an adventitious carbonaceous layer.  This is presumed to be a hydrocarbon layer, since the work function ($\sim$ 4.3 eV) and the C 1s core-level binding energy ($\sim$ 284.9 eV), shown in figure 4(b), are consistent with hydrocarbon values in the literature \cite{Tarlov1992, NIST2000} and are significantly different from that of atomic carbon on Au \cite{Jooya2018}.  After the first sputter treatment to the trap electrode, the work function and C 1s binding energy approach the values of the model system, C/Au(110).  Measurements after subsequent treatments that stepwise decrease the coverage, continue to resemble the model system. These data are consistent with the initial hydrocarbon layer being decomposed by the first sputter treatment that leaves a carbon residue on the electroplated Au surface \cite{Contarini1987,Marletta1992,Toth1994,Iijima1994,Lesiak1995,Koizumi2001}.  More work is needed to further verify this, however it provides a clue about how the electric-field noise above an electrode surface can depend on the composition and coverage of a contamination layer.

To study this idea in more detail, spatial distributions from KPFM images were examined and likewise compared to the model system, shown in figure 5.  This provides additional information, since the work functions discussed previously are average values of CPD measurements over a given area (300 nm $\times$ 300 nm).  Due to the patchy nature of the electroplated Au, the red circle data points in figure 4(a) should actually be smeared out areas or regions of distributed work functions.  The coverage measurement is an average over a 200 $\mu$m diameter analysis area with no spatial resolution.

\begin{figure}
\centering
\includegraphics[width=0.8 \textwidth, keepaspectratio]{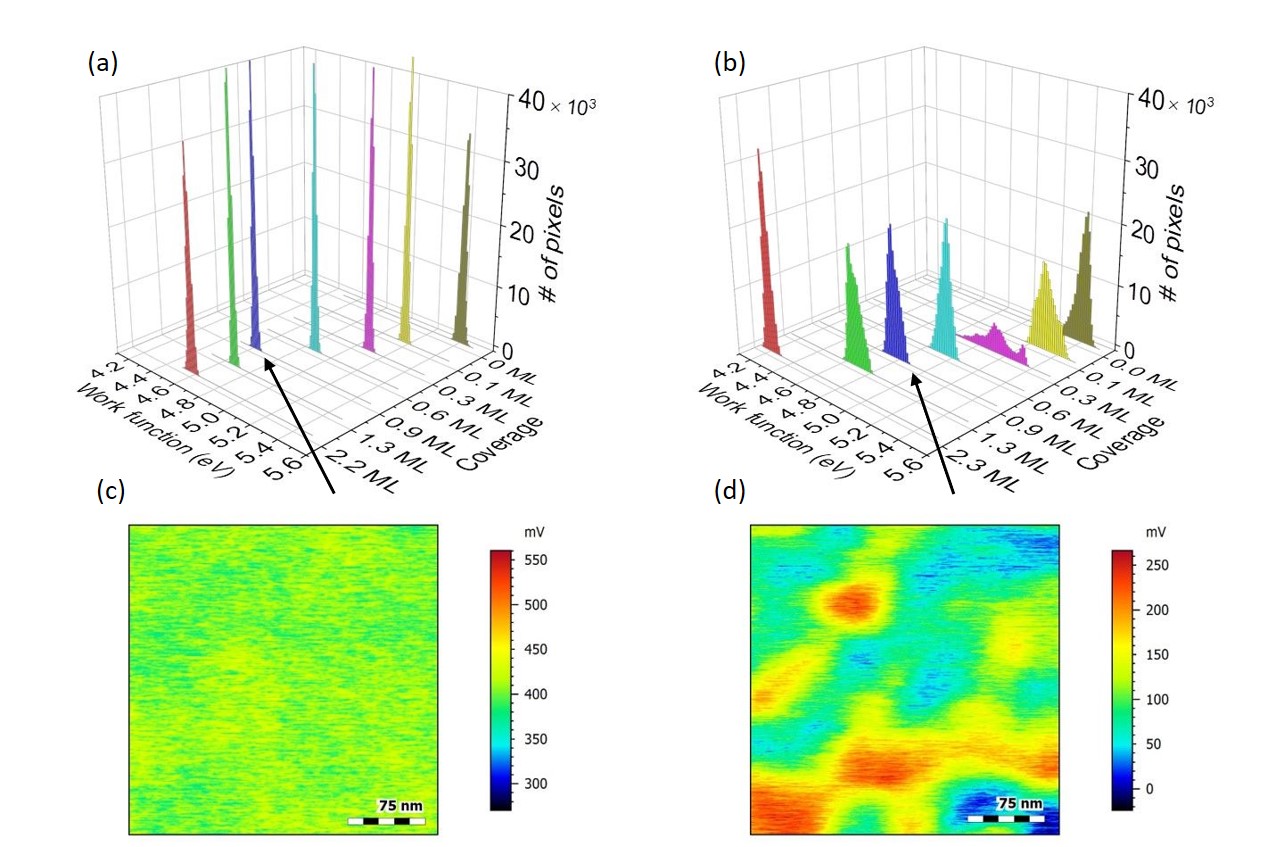}
\caption{Histograms of KPFM images for various coverages for (a) C/Au(110) and (b) the sample ion-trap electrode.  C/Au(110) exhibits narrow work-function distributions. The sample ion-trap electrode shows broad, multi-peak distributions, which indicate much greater surface-potential ``roughness" and more work-function states for fluctuations.  Representative KPFM CPD images at 0.9 ML for (c) C/Au(110) and (d) the sample electrode show the patchy nature of the surface potentials.  Images are of 300 nm $\times$ 300 nm areas and are plotted on the same (relative) color scale.}
\label{fig:histograms}
\end{figure}

\section{Discussion}

In many analyses of electric-field noise from surfaces, models of work-function patches or adsorbate dipoles are considered to be on atomically flat electrodes.  In actuality, metal electrodes tend to be rough, particularly electroplated electrodes with a typical RMS roughness on the order of 10s of nanometers. In \cite{Lin2016}, the effects of surface roughness on electric-field noise was investigated theoretically.  It was found that, at low temperatures, heating rates are enhanced or suppressed depending on the adatom coverage in morphological regions of negative or positive curvature, respectively \cite{Lin2016}.  

Here, the roughness of the surface potential is considered, rather than the topographical roughness, to describe a landscape for mobile dipoles to cause the fluctuation of work-function patches giving rise to electric-field noise.  In this work, patch dimensions on the sample ion-trap electrode were 50 nm to 100 nm, shown in figure 5(d).  Since the work function of a given patch depends on the adsorbate coverage, particularly in the submonolayer regime (see figure 4), temporal changes in the local coverage give rise to work-function fluctuations.  

The mechanism to drive such adsorbate motion and cause coverage fluctuations remains unclear.  Moreover, the physical fluctuations need to be fast to correspond to the high-frequency noise measured in ion traps.  It was suggested in \cite{Hite2017} that fields or currents from the trap RF drive may play a role in adsorbate fluctuation.  In those experiments, heating from samples positioned near a trapped ion (varied between 45 $\mu$m and 300 $\mu$m) was reasoned to be negligible compared to the low heating rates from the sputter-treated trap electrodes.  It was suggested that, due to the large impedance of the sample to RF ground, currents and fields may have been too low at the samples' surface.  In a similar, more recent experiment with modified sample wiring to lower the impedance at the RF-drive frequency, heating from nearby samples was observed, however tests with additional RF pulses applied to the samples showed no increase to the measured heating rates \cite{McKay2021}.  A significant challenge to theory and experiment alike would be to develop and test a model that relates adatom fluctuations driven at the trap-drive frequency resulting in motion and electric-field noise at the lower trapped-ion motional frequencies. This may involve considerations of complex adatom-adatom, adatom-substrate and/or cluster interactions at the surface.  This is not well understood and warrants further theoretical and experimental investigation.

To illustrate the surface-potential roughness, histograms of KPFM images as a function of coverage are shown in figure 5 for (a) the model system C/Au(110) and (b) the sample ion-trap electrode.  The width of these histograms represents this roughness and the variation of work functions over a given area.  On the one hand, as shown in figure 5(a), C/Au(110), prepared by carbon deposition, exhibits narrow distributions at or below the resolution of the measurement for all coverages shown.  These are histograms of KPFM images collected for the work-function experiments described in \cite{Jooya2018}.  On the other hand, the electroplated Au electrode, prepared by sputter etching, exhibits broad and multipeak distributions, indicating a patchy surface whose widths change nonmonotonically with decreasing coverage (figure 5(b)).   Figs. 5(c, d) show representative KPFM CPD images for 0.9 ML of C/Au(110) and the treated sample ion-trap electrode, respectively (same relative color scale).

An electrode with more patches will have a greater work-function noise spectral density, proportional to the surface-potential roughness, for a given degree of coverage fluctuations.  To relate this to the nonmonotonic heating-rate measurements shown in figure 2(a), a simple empirical model, dependent on the adsorbate coverage and the surface-potential roughness, was used to arrive at a heating-rate trend for the sample ion-trap electrode.  The model assumes a dependence on the coverage $\theta$ such that the heating rate will tend to zero with decreasing coverage and at higher coverages will asymptotically approach a maximum value with a characteristic decay constant $\lambda$, such as [1-exp(-$\theta/\lambda$)]. Ideally, as the coverage is reduced through the low coverage regime ($\theta \leq$ 0.5 ML), the global work function will approach that of the clean electrode and coverage fluctuations and hence work function fluctuation will become more dilute with an approximately linear behavior.  As the coverage increases through the high coverage regime ($\theta >$ 0.5 ML), the work function will approach and reach that of the adsorbate overlayer, like a bulk material, and eventually the noise will cease to increase with additional adsorbate layers.

The RMS surface-potential roughness $\sigma_{\phi}$, corresponding to the standard deviation of a CPD image with N pixels per row and M pixels per column, is given by
\begin{equation}
    \sigma_{\phi}= \left(\frac{1}{A}\sum_{m=1}^{M}\sum_{n=1}^{N}(\phi_{m,n}-\bar{\phi})^2\right)^{1/2},
\end{equation}
where ${\phi}$ $(= eV_{CPD})$ is the work function and \textit{A} ($=M{N}\cdot$[area/pixel]) is the area of interest.  If one assumes a first order dependence on $\sigma_{\phi}$, the empirical heating-rate trend is given by
\begin{equation}
    \dot{\bar{n}} = C\cdot(1-e^{-\theta/\lambda})\cdot\sigma_{\phi}+bkg,
\end{equation}
where a prefactor \textit{C}, $\lambda$ and an additional background offset (\textit{bkg}) are found from a pointwise least-squares fit to the heating-rate data.  A Gaussian function, which fits well to $\sigma_{\phi}(\theta)$ for the sample electrode, was used to approximate the unknown  $\sigma_{\phi}(\theta)$ for the stylus-trap electrodes.  A modelled $\dot{\bar{n}}$ can be determined for qualitative purposes by determining $\sigma_{\phi}$ from the KPFM images for each coverage $\theta$. The data from (3) for the sample electrode, shown in figure 6, have a \textit{C} = 4 (quanta per second)/meV, \textit{bkg} = 4 quanta per second, and $\lambda$ = 0.2 ML, showing the relevance of the submonolayer regime to ion heating, previously pointed out in \cite{Daniilidis2014}. For example, from the sample ion-trap electrode shown in figure 5(d) with $\theta$ = 0.9 ML and $\sigma_{\phi}$ = 42 meV, equation (3) gives 170 quanta per second, while the C/Au(110) sample, with $\theta$ = 0.9 ML and $\sigma_{\phi}$ = 11 meV in figure 5(c), has a value of 48 quanta per second. 

\begin{figure}
\centering
\includegraphics[width=0.6 \textwidth, keepaspectratio]{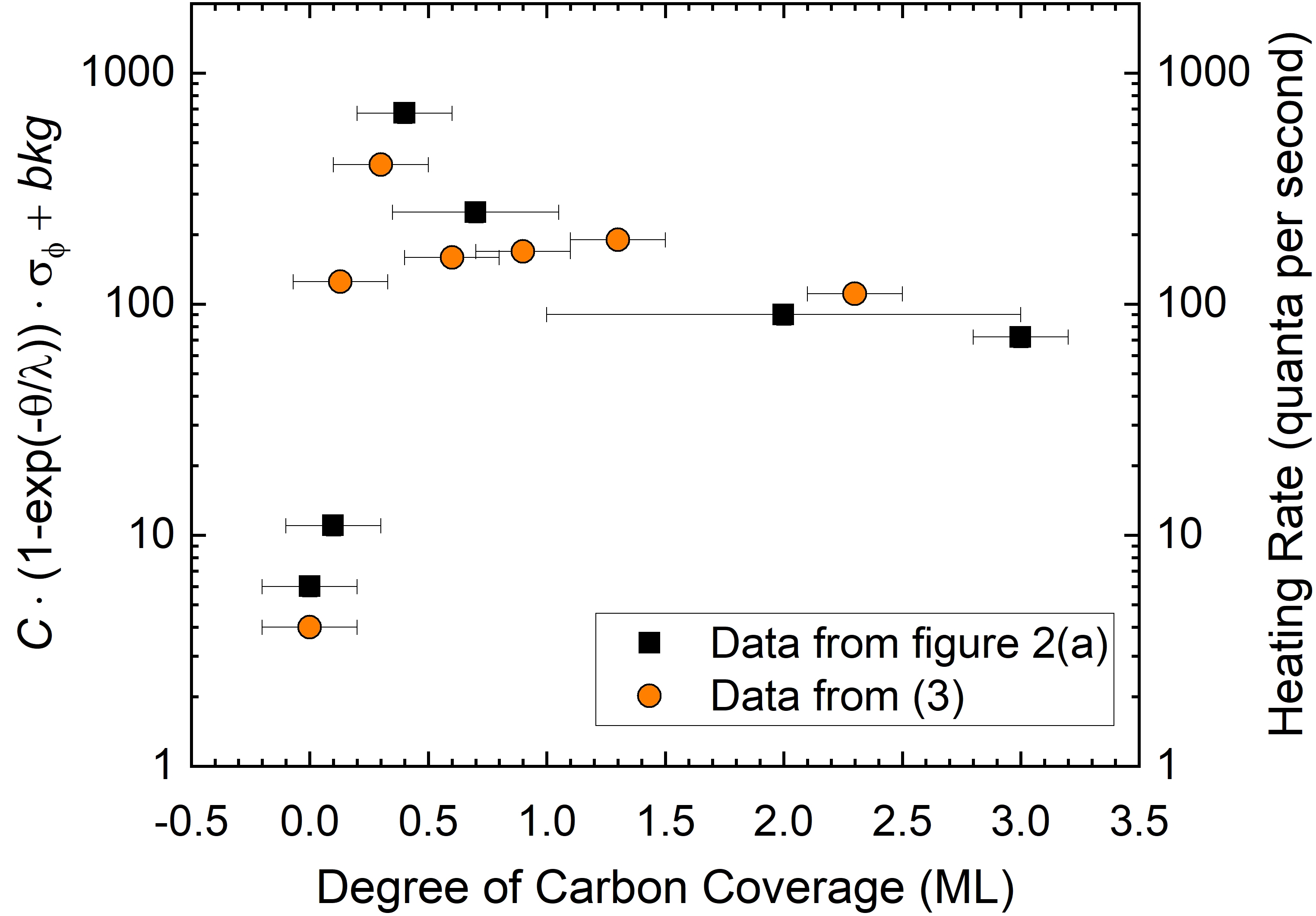}
\caption{Plot of the modelled heating rate from (3) using the measured values for the coverage $\theta$ and the surface-potential roughness $\sigma_{\phi}$, where $\phi$ is the work function, $C$ is a scaling constant, and \textit{bkg} is an additional background offset.  The trend qualitatively agrees with the heating rate data from figure 2(a).}
\label{fig:model}
\end{figure}

 Additional experimental work is required to corroborate this model with variations in trapped-ion heating rates, but it gives a qualitative explanation for the nonmonotonic heating observed in ion traps with electrodes of various adsorbate coverages.  Due to variability in ion-trap electrode preparation, the metric of surface-potential roughness could help to understand the spread in heating rates reported throughout the literature over the past few decades \cite{Brownnutt2015}.  A future study of KPFM and ion heating rates measured with the same sample surface will allow for a quantitative analysis of this and other models.  In addition, this simple model suggests that clean electrode surfaces or even an adsorbate covered surface with a uniform surface-potential landscape may enable ion-trap systems with low heating rates aiding QIP scalability through miniaturization.

\section{Summary}

In this article, the frequency scaling of the electric-field noise in an incrementally sputter-treated ion trap was shown to change from a power-law exponent $\alpha$ of $\sim$ 1.5 to $\sim$ 0.4. This variation is correlated with the nonmonotonic behavior in the heating rate as a function of carbon coverage on the electroplated Au trap electrodes.  These values of $\alpha$ are consistent with some aspects of adsorbate-dipole models due to the effects of surface treatments or changes in coverage.  In addition, an anisotropy in the heating rates of the two orthogonal motional modes (parallel to the electrode surfaces) was shown to become more pronounced in the low-coverage regime.  At a motional frequency of 3 MHz, the mode direction in line with the sputter-beam direction experienced $\sim$ 3.5 times higher electric-field noise than the mode perpendicular to the sputter beam after four sputter treatments, possibly pointing to a morphological role in the heating from the surface.

To elucidate these results, a KPFM/XPS study was conducted on a sample ion-trap electrode, comparing the coverage dependent work function to that of a model system, atomic carbon on Au(110).  The as-fabricated sample electrode was covered with a hydrocarbon film with very different properties from C/Au(110). Even though, after incremental sputter treatments, the sample electrode resembled the average work functions and C 1s binding energies of the model system, in contrast to deposited carbon on Au(110), KPFM revealed fundamentally different surface-potential maps with work-function patch dimensions $\sim$ 50 nm to 100 nm.  Finally, measured coverage and surface-potential roughness were employed in a simple empirical model to qualitatively connect the morphology of local work functions to the nonmonotonic heating rates observed from incrementally treated ion-trap electrodes.    

\ack

The authors acknowledge the following for many fruitful discussions surrounding this work: D. Leibfried, S. Kotler, H. R. Sadeghpour, A. C. Wilson, D. J. Wineland, and especially D. H. Slichter for also fabricating and providing the sample ion-trap electrodes.  The authors also thank A. P. McFadden, T. M. Wallis, and R. B. Goldfarb for helpful comments on the manuscript. 
K.S.M. acknowledges support  as an associate in the Professional Research Experience Program (PREP) operated jointly by the National Institute of Standards and Technology (NIST) and the University of Colorado Boulder under Award No. 70NANB18H006 from the U.S. Department of Commerce, NIST.

\section*{References}

\bibliographystyle{unsrt}
\bibliography{references}

\end{document}